\documentclass[12pt,a4paper,aps,prd,preprint,superscriptaddress,nofootinbib]{revtex4-1}
\usepackage[utf8]{inputenc}
\usepackage{graphicx}
\usepackage{amssymb}
\usepackage{textcomp}
\usepackage{amsmath}
\usepackage{tabularx}
\usepackage{bm}
\usepackage{times}
\usepackage{color}
\usepackage{slashed}
\usepackage{multirow}
\usepackage{verbatim}
\usepackage{cancel}
\usepackage{subfigure}
\usepackage[normalem]{ulem}

\usepackage[colorlinks=true, pdfstartview=FitV, linkcolor=blue, citecolor=blue, urlcolor=blue]{hyperref}
\allowdisplaybreaks[4]

\def\lsim{\mathrel{\raise.3ex\hbox{$<$\kern-.75em\lower1ex\hbox{$\sim$}}}}
\def\gsim{\mathrel{\raise.3ex\hbox{$>$\kern-.75em\lower1ex\hbox{$\sim$}}}}

\newcommand{\bew}{\begin{widetext}}
\newcommand{\enw}{\end{widetext}}
\newcommand{\bee}{\begin{equation}}
\newcommand{\ene}{\end{equation}}
\newcommand{\bea}{\begin{eqnarray}}
\newcommand{\ena}{\end{eqnarray}}
\newcommand{\bes}{\begin{subequations}}
\newcommand{\ens}{\end{subequations}}

\def\calb{\mathcal{B}}

\def\calm{\mathcal{M}}

\def\calr{\mathcal{R}}

\definecolor{orange}{rgb}{1,0.5,0}

\begin{document}

\title{Laser induced Compton Scattering to Dark Photon or Axion-like Particle}

\author{Kai Ma}
\email{makai@ucas.ac.cn}
\affiliation{Faculty of Science, Xi'an University of Architecture and Technology, Xi'an, 710055, China}

\author{Tong Li}
\email{litong@nankai.edu.cn}
\affiliation{
School of Physics, Nankai University, Tianjin 300071, China
}

\begin{abstract}
The laser of an intense electromagnetic field provides an important tool to study the strong-field particle physics. The nonlinear Compton scattering was observed in the collision of
an ultra-relativistic electron beam with a laser pulse in 1990s. The precision measurement of the nonlinear Compton scattering shines the light on the studies of strong-field QED and nonlinear effects in QED. In this work, we propose to produce new massive particles through the nonlinear Compton scattering in the presence of a high intense laser field. We take dark photon or axion-like particle as illustrative new particle.
Compared with other collider and beam dump experiments, the laser induced process provides a complementary and competitive search of new invisible particles lighter than 1 MeV.
\end{abstract}

\maketitle
\tableofcontents

\section{Introduction}
\label{sec:Intro}

The laser of an intense electromagnetic field strength has a lot of applications in atomic physics, nuclear physics and strong-field particle physics (see Greiner et al.'s textbook~\cite{Greiner:1992bv} and a recent review~\cite{Fedotov:2022ely}). For instance, the first laser
excitation of the Th-229 low-energy nuclear transition has recently shined the light on a nuclear clock~\cite{Tiedau:2024obk}. A proton can decay into a neutron, positron and electron-neutrino in the presence of a strong electromagnetic field~\cite{Wistisen:2020czu,Ouhammou:2022jys}. It can change the properties of elementary particles or induce processes that cannot occur in vacuum. Thus, the intense laser pulse provides an essential tool in exploring the high-intensity frontier of physics.

In 1990s, the experiment performed at SLAC observed two strong-field processes in the interaction of
an ultra-relativistic electron beam with a terawatt laser pulse, i.e. the nonlinear Compton scattering and the electron-positron pair production in high intense laser field~\cite{Burke:1997ew,Bamber:1999zt}. The precision measurements of these two processes enable the studies of strong-field quantum electrodynamics (QED) and nonlinear effects in QED (see a recent review in Ref.~\cite{Hartin:2018egj}). In the nonlinear Compton scattering, an electron beam absorbs multiple photons from the laser field and radiates a single photon
\begin{eqnarray}
e^- + {\rm laser} \to e^- + \gamma\;.
\end{eqnarray}
The nonlinear QED effect is characterized by the simultaneous interaction of electron with a large number of optical photons. When the strength of the electromagnetic plane wave is high enough, the nonlinear effects from the higher-order terms in the scattering cross section become important. More precisely, the nonlinear effects are significant when
\begin{eqnarray}
{e\varepsilon_0\over \omega m_e} > 1\;,
\end{eqnarray}
where $e$ is the electric charge unit, $\varepsilon_0$ is the electromagnetic field strength, $m_e$ is the electron mass and $\omega$ is the laser beam energy. The case of more than one initial photon corresponds to higher-multipole radiation. The accumulation arising from the absorption of more than one photon results in a sizable cross section in unit of barn.
This process was also utilized to manufacture a high-energy photon for $\gamma\gamma$ collision. A recent experiment to explore the behavior of QED in the strong-field regime is the LUXE at DESY~\cite{LUXE:2023crk}.

In this work, instead of an energetic photon, we propose to produce new particles beyond the Standard Model (SM) in this nonlinear Compton scattering. We take two intriguing invisible particles as illustration, i.e. dark photon (DP) and axion-like particle (ALP). The search for DP or ALP with mass from several eVs to MeVs is subject to collider experiments or beam dump experiments. The beam dump experiments are useful for exploring the regime of small coupling but lose sensitivity for mass lower than $2m_e\sim 1$ MeV. The collider experiments are able to probe smaller mass region but suffer from low production rate and available data. In our case, a high-energy electron beam collides with an intense laser beam of a number of optical photons to form a massive new particle
\begin{eqnarray}
e^- + {\rm laser} \to e^- + {\rm DP~or~ALP}\;.
\end{eqnarray}
The decay $e^-\to e^- + {\rm massive~particle}$ is forbidden in vacuum but can occur after absorbing optical photons. The new particle couples to electron and is assumed to be long-lived to escape from the detector. The sizable cross section of this process for the light new particle creation leads to an increased observation sensitivity compared to other collider experiments or beam dump experiments.
The idea of nonlinear Compton scattering to ALPs was originally proposed in Refs.~\cite{Dillon:2018ypt,King:2018qbq}. The energetic photon from Compton scattering can be formed as an optical dump experiment to search for ALP~\cite{Bai:2021gbm}. Ref.~\cite{Dillon:2018ouq} also proposed to detect the produced ALP through the Light-Shining-through-Wall (LSW) setup. Other related studies include Refs.~\cite{King:2019cpj,Beyer:2021mzq}. The main original contribution of our work is in the study of DPs.

This paper is organized as follows. In Sec.~\ref{sec:LaserCompton}, we review the Compton scattering in intense laser field. The strong-field QED is discussed in details. In Sec.~\ref{sec:DPALP}, we discuss the laser-induced Compton scattering to invisible particles. The DP and ALP are taken as illustrative examples of invisible particle. We present the cross sections of the relevant Compton scattering and show the sensitivity reach for model parameters in Sec.~\ref{sec:results}. Our conclusions are drawn in Sec.~\ref{sec:Con}. The details of Lorentz transformation for our calculation are collected in Appendix~\ref{app:rotation}.

\section{Laser induced Compton scattering}
\label{sec:LaserCompton}

We start from reviewing the following Compton scattering process in intense laser field
\begin{eqnarray}
e^-(p) + n\omega (k) \to e^-(p') + \gamma (k')\;,
\end{eqnarray}
where $n$ denotes the number of optical photons with energy $\omega$. The details were well presented in Greiner et al.'s QED textbook~\cite{Greiner:1992bv}.

In the presence of an electromagnetic potential, the wave function of a relativistic fermion of mass $m$ is subjected to the following Dirac equation
\bee
(i \slashed{\partial} - Qe \slashed{A} -m) \psi(x)=0 \;,
\ene
where $e$ is the unit of electric charge and $Q$ is the charge operator (e.g. $Q\psi=-\psi$ for electron).
With circular polarization, the vector potential $A^\mu$ has the following form
\bee
\label{eq:laser}
A^\mu(\phi)= a_1^\mu \cos (\phi)+ a_2^\mu \sin (\phi) \,,\quad \phi=k \cdot x,
\ene
where $k$ is the electromagnetic wave vector and
$\phi$ is the phase of the laser field. The polarization vectors are given as
\bea
a_1^\mu = |\vec{a}\,|(0,1,0,0)\,,~~~
a_2^\mu = |\vec{a}\,|(0,0,1,0)\;.
\ena
They satisfy the following relations
\bea
a_1 \cdot a_2 = 0\;,~~~
a_1^2 = a_2^2 = -|\vec{a}\,|^2 \equiv -a^2\;,
\ena
where $a=\varepsilon_0/\omega$ with $\varepsilon_0$ being the strength of the electric and magnetic field.
Note that here we choose the laser background to be circularly polarized and monochromatic~\cite{King:2018qbq}. In a realistic experimental setup, the high-intensity laser would emit the photons in pulses. One can introduce a function $f(x)$ for the pulse shape describing the spatial dependence of the vector potential in the parameterization of $A^\mu$~\cite{Dillon:2018ypt}. The pulse shape for $f(x)$ is supposed to be Gaussian with respect to the phase $\phi$. The spatial dependence of the external field can enter the following calculation of the $S$ matrix element through the Fourier transform of the pulse shape $\tilde{f}$. A more complicated expression can then be obtained. We instead work in the monochromatic limit in our calculation.
The lowest-order scattering $S$ matrix element for the laser-induced production reads
\begin{eqnarray}
S_{f i}
&=&
ie {1\over \sqrt{2k^{\prime 0} V}}\int d^4 x e^{ik'\cdot x}\; \overline{\psi_{p', s'}(x)} \cancel{\varepsilon} \psi_{p, s}(x) \;,
\end{eqnarray}
where $\varepsilon^\mu$ is the polarization vector of the outgoing physical photon.
The wave function of the incoming electron with $Q=-1$ is given by the Volkov state~\cite{Wolkow1935}
normalized to the volume $V$ as follows
\bee
\psi_{p, s}(x)
=
\left[1+\frac{Qe \slashed{k} \slashed{A}}{2\left(k \cdot p\right)}\right]
\frac{u\left(p, s\right)}{\sqrt{2 q^0 V}} e^{i F_1(q, s)},
\ene
where $u(p, s)$ is the usual Dirac spinor for the free electron. It gains an ``effective momentum''
\bee
q^\mu
=
p^\mu + \frac{Q^2 e^2 a^2}{2 k \cdot p} k^\mu
\ene
which satisfies the dispersion relation
\bee
q^2 = m^2_e + Q^2 e^2 a^2 = m_e^{\ast 2} \,.
\ene
The phase function $F_1(q, s)$ is given by
\bee
F_1(q, s)
=
-q \cdot x - \frac{Qe\left(a_1 \cdot p\right)}{\left(k \cdot p\right)} \sin \phi
+
\frac{Qe\left(a_2 \cdot p\right)}{\left(k \cdot p\right)} \cos \phi \;.
\ene
Similarly, the Volkov state of the final electron in the laser field can be obtained with the substitution $p\to p'$ and $q\to q'$. The Volkov states are the exact solutions of the Dirac equation describing the ``dressed'' electrons which continuously
interact with the travelling cloud of laser photons. They automatically include the nonlinear effects of the laser field in the series of $ea/m_e$.

The scattering amplitude is
\begin{eqnarray}
\label{eq:am}
\calm
&=&ie {1\over \sqrt{2k^{\prime 0} V}}e^{ik'\cdot x}\overline{\psi_{p', s'}(x)} \cancel{\varepsilon} \psi_{p, s}(x)\nonumber\\
&=&
ie {e^{i(k'+q'-q)\cdot x}e^{-i\Phi}\over \sqrt{2^3 V^3 q^0 q^{\prime 0} k^{\prime 0}}} \overline{u(p', s')} \left[1-\frac{e \slashed{A}\slashed{k}  }{2 k \cdot p'}\right]\cancel{\varepsilon} \left[1-\frac{e \slashed{k} \slashed{A} }{2 k \cdot p}\right] u(p,s)\;,
\end{eqnarray}
where the phase factor $\Phi$ is given by
\bee
\Phi =
e a_1 \cdot y \sin \phi - e a_2 \cdot y \cos \phi
\ene
with $y$ being defined as
\bee
y^\mu = \frac{ p^{\prime \mu} }{k \cdot p'} - \frac{ p^\mu }{k \cdot p}\;.
\ene
Substituting the photon field given in \eqref{eq:laser}, the amplitude can be factorized as follows
\begin{eqnarray}
\calm
&=&
ie {e^{i(k'+q'-q)\cdot x}\over \sqrt{2^3 V^3 q^0 q^{\prime 0} k^{\prime 0}}}
\sum_{i=0}^{2}
C_i \calm_i\;.
\end{eqnarray}
The reduced amplitudes $\calm_i$ are
\bea
\calm_0
&=&
\overline{u_2} \cancel{\varepsilon} u_1 +\overline{u_2}{e^2 a^2\over 2k\cdot p k\cdot p'} k\cdot \varepsilon \cancel{k}u_1\;,
\\[3mm]
\calm_1
&=&
-\overline{u_2}\cancel{\varepsilon}
\frac{e \slashed{k} \slashed{a}_1 }{2 k \cdot p} u_1 - \overline{u_2}
\frac{e \slashed{a}_1 \slashed{k} }{2 k \cdot p'} \cancel{\varepsilon} u_1\;,
\\[3mm]
\calm_2
&=&
-\overline{u_2}\cancel{\varepsilon}
\frac{e \slashed{k} \slashed{a}_2 }{2 k \cdot p} u_1 - \overline{u_2}
\frac{e \slashed{a}_2 \slashed{k} }{2 k \cdot p'} \cancel{\varepsilon} u_1\;.
\ena
The corresponding coefficients are given by
\bea
C_0 &=& e^{-i\Phi}\;,
\\[3mm]
C_1 &=& \cos\phi \,e^{-i\Phi}\;,
\\[3mm]
C_2 &=& \sin\phi \,e^{-i\Phi}\;,
\ena
where the phase $\Phi$ takes the form
\bee
\Phi  =  z \sin(\phi - \phi_0)
\ene
with
\bee
z = e \sqrt{ (a_1 \cdot y)^2 + (a_2\cdot y)^2 } \,,
\quad
\cos\phi_0 =  \frac{e a_1 \cdot y }{ z} \,,
\quad
\sin\phi_0 =  \frac{e a_2 \cdot y}{ z} \,.
\ene
Sine the function $e^{-i\Phi}$ is a periodic function, it can be expanded as
\bee
e^{-i\Phi}
=
e^{- i z \sin \left(\phi-\phi_0\right)}
=
\sum_{n=-\infty}^{\infty} c_n e^{-\mathrm{i} n\left(\phi-\phi_0\right)}\;,
\ene
where the coefficients are given as
\bee
c_n
=
\frac{1}{2 \pi} \int_{-\pi}^\pi \mathrm{d} \varphi \, e^{-\mathrm{i} z \sin\varphi} e^{\mathrm{i} n \varphi}
=J_n(z)\;.
\ene
This is the integral representation of the Bessel function $J_n(z)$. Thus, we have
\bee
C_0=e^{-i\Phi} = \sum_{n=-\infty}^{\infty} B_n(z) e^{-\mathrm{i} n \phi}\;,
\ene
where $B_n(z)=J_n(z) \mathrm{e}^{i n \phi_0}$. This is a well-known relation known as Jacobi-Anger expansion.
By using the following relations
\bea
\cos\phi = \frac{1}{2} \Big( e^{i\phi} + e^{-i\phi} \Big) \,,~~~
\sin\phi = \frac{1}{2i} \Big( e^{i\phi} - e^{-i\phi} \Big) \,,
\ena
we have
\bea
C_1=
\cos\phi \,e^{-i\Phi}
&=&
\frac{1}{2}\sum_{n=-\infty}^{\infty} B_n(z)  \Big[ e^{-i(n-1)\phi} + e^{-i(n+1)\phi} \Big]\;,
\\[3mm]
C_2=
\sin\phi \,e^{-i\varPhi}
&=&
\frac{1}{2i}\sum_{n=-\infty}^{\infty} B_n(z)  \Big[ e^{-i(n-1)\phi} - e^{-i(n+1)\phi} \Big]\;.
\ena
By redefining the integer $n' = n+1$ and  $n'' = n-1$, together with the completeness of the $n'$ and $n''$, we obtain
\bea
C_1=
\cos\phi \,e^{-i\varPhi}
&=&
\frac{1}{2}\sum_{n=-\infty}^{\infty} \Big[ B_{n+1}(z) + B_{n-1}(z)  \Big] e^{-in\phi} \;,
\\[3mm]
C_2=
\sin\phi \,e^{-i\varPhi}
&=&
\frac{1}{2i}\sum_{n=-\infty}^{\infty}  \Big[ B_{n+1}(z) - B_{n-1}(z)  \Big] e^{-in\phi} \;.
\ena

Then, the amplitude can be further simplified as
\bee
\calm
=
ie \sum_{n=-\infty}^{\infty}
\frac{ e^{i( k' +q' - q - nk) \cdot x }  }{ \sqrt{2^3 V^3 q^0 q^{\prime 0} k^{\prime 0} }}
\sum_{i=0}^{2}
\widetilde{C}_i^{n} \calm_i\;,
\ene
where
\bee
\widetilde{C}_i^{n} = e^{in\phi} C_i^{n} \;,~~~\sum_{n=-\infty}^{\infty}\widetilde{C}_i^{n} e^{-in\phi}=\sum_{n=-\infty}^{\infty}C_i^n=C_i\;.
\ene
After integrating out the coordinate $x$, we have
\bee
S_{f i}
=
ie \sum_{n=-\infty}^{\infty}
\frac{ (2\pi)^4 \delta^4( k' + q' - q - nk) }{ \sqrt{2^3 V^3 q^0 q^{\prime 0} k^{\prime 0} }}
\sum_{i=0}^{2}
\widetilde{C}_i^{n} \calm_i\;,
\ene
and its square is given by
\bee
\big| S_{f i}\big|^2
=
e^2 \sum_{n=-\infty}^{\infty}
\frac{ (2\pi)^4 \delta^4( k' + q' - q - nk) VT }{ 2^3 V^3 q^0 q^{\prime 0} k^{\prime 0} }
\sum_{i,j=0}^{2}
\widetilde{C}_i^{n} \big( \widetilde{C}_j^{n} \big)^\dag  \overline{ \calm_i \calm_j^\dag }\;.
\ene
The ``overline'' in above equation means the summation and average over the spin degree of freedom of relevant particles. For the Compton scattering, after averaging the electron spin, the squared amplitude is then
\begin{eqnarray}
\sum_{i,j=0}^{2}
\widetilde{C}_i^{n} \big( \widetilde{C}_j^{n} \big)^\ast  \overline{ \calm_i \calm_j^\dag }&=&-4m_e^2 J_n^2(z)+e^2a^2{1+u^2\over u} [J_{n-1}^2(z)+J_{n+1}^2(z)-2J_n^2(z)]\;,
\end{eqnarray}
where $u\equiv k\cdot p/k\cdot p'$. We find
\begin{eqnarray}
z={2nu ea \over s_n'-m_e^{\ast 2}} \Big({s_n'+m_e^{\ast 2}-m_\chi^2\over u}-{s_n'\over u^2}-m_e^{\ast 2}\Big)^{1/2}\;.
\end{eqnarray}
The cross section of the Compton scattering is given by~\cite{Greiner:1992bv}
\begin{eqnarray}
\sigma &=& {|S_{fi}|^2\over VT}{1\over 2(1/V)}{1\over \rho_\omega} V \int \frac{d^3 q'}{(2 \pi)^3} V \int \frac{d^3 k'}{(2 \pi)^3}\nonumber\\
&=&{1\over 2\rho_\omega}\frac{ e^2 }{2 q^0 } \sum_{n=-\infty}^{\infty}  \int d\Pi_2
\sum_{i,j=0}^{2}
\widetilde{C}_i^{n} \big( \widetilde{C}_j^{n} \big)^\dag
\overline{\calm_i \calm_j^\dag} \;,
\label{eq:xsec}
\end{eqnarray}
where $\rho_\omega={a^2 \omega\over 4\pi}$ is the laser photon density~\cite{Greiner:1992bv}, and $d\Pi_2$ is the standard two-body phase space
\bee
d\Pi_2
=
\int \frac{d^3 q'}{(2 \pi)^3 2 q^{\prime 0} }
\int \frac{d^3 k'}{(2 \pi)^3 2 k^{\prime 0} }
(2\pi)^4 \delta^4( k' + q' - q - nk) \;.
\ene
In the center-of-mass (c.m.) frame with c.m. energy $\sqrt{s_n'}$ and $s_n'\equiv (q+nk)^2$, the cross section is
\begin{eqnarray}
\sigma &=&{e^2 \over 32\pi q^0 \rho_\omega}\sum_{n=-\infty}^{\infty} \sum_{i,j=0}^{2}
\widetilde{C}_i^{n} \big( \widetilde{C}_j^{n} \big)^\ast  \overline{ \calm_i \calm_j^\dag }  
\int  du {1\over u^2}\;,
\end{eqnarray}
where
\begin{eqnarray}
{\sqrt{s_n'}\over Q'+|\vec{q}'|}<u<{\sqrt{s_n'}\over Q'-|\vec{q}'|}
\end{eqnarray}
with
\begin{eqnarray}
Q'={s_n'+m_e^{\ast 2}-m_\chi^2\over 2\sqrt{s_n'}}\;,~~~
|\vec{q}'|={\lambda^{1/2}(s_n',m_e^{\ast 2},m_\chi^2)\over 2\sqrt{s_n'}}\;,
\end{eqnarray}
and $\lambda(x,y,z)=x^2+y^2+z^2-2xy-2xz-2yz$.
Note that here we keep the photon mass denoted by $m_\chi$ for the convenience of the calculation of the scattering to massive particles in next section.
In the realistic collisions of electrons with intense laser pulses, the laser beam crosses the electron beam at a nonzero angle $\theta_{\rm Lab}$ (e.g. $17^\circ$ at SLAC)~\cite{Bamber:1999zt}. To show the kinematic distributions, we need to rotate and boost the c.m. frame to the laboratory frame. The details of rotation and boost are given in Appendix~\ref{app:rotation}.

\section{Compton scattering to new massive particles in intense laser field}
\label{sec:DPALP}

In this section, we apply the above method of laser induced Compton scattering and investigate the scattering to new massive particles in intense laser field. As illustrative examples, we consider dark photon $\gamma_D$ and ALP $a$ as new massive particles in the following Compton scattering processes
\begin{eqnarray}
e^-(p) + n\omega (k) \to e^-(p') + \gamma_D/a (k')\;.
\end{eqnarray}

The dark photon (or called hidden photon)~\cite{Holdom:1985ag,Holdom:1986eq} is a spin-one particle gauged by an Abelian group $U(1)_{\rm D}$ in dark sector (see a recent review Ref.~\cite{Fabbrichesi:2020wbt} and references therein). The interaction between the visible photon and the dark photon is given by the gauge kinetic
mixing between the field strength tensors of the SM electromagnetic gauge group $U(1)_{\rm EM}$ and the $U(1)_{\rm D}$ group. The DP Lagrangian reads as
\begin{eqnarray}
\mathcal{L}_{\rm DP}\supset -\frac{1}{4}F^{\mu\nu}F_{\mu\nu}-\frac{1}{4}{F}_D^{\mu\nu}{F}_{D\mu\nu}-\frac{\epsilon}{2}F^{\mu\nu}{F}_{D\mu\nu}+\frac{1}{2}m_{D}^2{A_D}^\mu A_{D\mu}\;,
\end{eqnarray}
where $F^{\mu\nu}$ ($F^{\mu\nu}_D$) is the SM (dark) field strength, and $A_D$ is the dark gauge boson with mass $m_{\gamma_D}$. Suppose the SM particles are uncharged under the dark gauge group, the kinetic mixing $\epsilon$ is received by integrating out heavy particles charged under both gauge groups at one-loop level. The two gauge fields can be rotated to get rid of the mixing. As a result, the SM matter current gains a shift by $A_\mu\to A_\mu - \epsilon A_{D\mu}$. Thus, the physical DP $\gamma_D$ has a direct coupling to the electromagnetic current of the SM particles
\begin{eqnarray}
\mathcal{L}_{\rm DP}\supset -e \epsilon J_{\rm EM}^\mu A_{D\mu}=-eQ\epsilon \overline{\psi}\gamma^\mu \psi A_{D\mu}\;.
\end{eqnarray}
The massive DP particle can then be searched in terrestrial experiments in the presence of this kinetic mixing. The amplitude of Compton scattering to DP is given by
\begin{eqnarray}
\calm^{\gamma_D}
&=&ie \epsilon {1\over \sqrt{2k^{\prime 0} V}}e^{ik'\cdot x}\overline{\psi_{p', s'}(x)} \cancel{\varepsilon}_D \psi_{p, s}(x)\nonumber\\
&=&
ie \epsilon {e^{i(k'+q'-q)\cdot x}e^{-i\Phi}\over \sqrt{2^3 V^3 q^0 q^{\prime 0} k^{\prime 0}}} \overline{u(p', s')} \left[1-\frac{e \slashed{A}\slashed{k}  }{2 k \cdot p'}\right]\cancel{\varepsilon}_D \left[1-\frac{e \slashed{k} \slashed{A} }{2 k \cdot p}\right] u(p,s)\;,
\end{eqnarray}
where $\varepsilon_D$ denotes the polarization vector of DP. The individual amplitudes $\calm_i^{\gamma_D}$ become
\bea
\calm_0^{\gamma_D}
&=&
\overline{u_2} \cancel{\varepsilon}_D u_1 +\overline{u_2}{e^2 a^2\over 2k\cdot p k\cdot p'} k\cdot \varepsilon_D \cancel{k}u_1\;,
\\[3mm]
\calm_1^{\gamma_D}
&=&
-\overline{u_2}\cancel{\varepsilon}_D
\frac{e \slashed{k} \slashed{a}_1 }{2 k \cdot p} u_1 - \overline{u_2}
\frac{e \slashed{a}_1 \slashed{k} }{2 k \cdot p'} \cancel{\varepsilon}_D u_1\;,
\\[3mm]
\calm_2^{\gamma_D}
&=&
-\overline{u_2}\cancel{\varepsilon}_D
\frac{e \slashed{k} \slashed{a}_2 }{2 k \cdot p} u_1 - \overline{u_2}
\frac{e \slashed{a}_2 \slashed{k} }{2 k \cdot p'} \cancel{\varepsilon}_D u_1\;.
\ena
After averaging the electron spin, the squared amplitude is then
\begin{eqnarray}
\sum_{i,j=0}^{2}
\widetilde{C}_i^{n} \big( \widetilde{C}_j^{n} \big)^\ast  \overline{ \calm_i^{\gamma_D} \calm_j^{\gamma_D\dag} }&=&-2(2m_e^2+m_\chi^2)J_n^2(z)\nonumber\\
&+&e^2a^2{1+u^2\over u} [J_{n-1}^2(z)+J_{n+1}^2(z)-2J_n^2(z)]\;,
\end{eqnarray}
where $m_\chi=m_{\gamma_D}$ and the maximal value of $m_\chi$ is given by $\sqrt{s_n'}-m_e^{\ast}$. The cross section of relevant Compton scattering is obtained by substituting $e^2$ in Eq.~(\ref{eq:xsec}) by $e^2\epsilon^2$.

The axion-like particle is a CP-odd pseudo-Nambu-Goldstone boson as a result of the spontaneous breaking of a global $U(1)$ symmetry. One well-studied example of ALP is the QCD axion~\cite{Peccei:1977hh,Peccei:1977ur,Weinberg:1977ma,Wilczek:1977pj} (see a recent review Ref.~\cite{DiLuzio:2020wdo}). The ALP can also appear in many other models in theory.
In general, the ALP mass ($m_a$) and the symmetry breaking scale (or called decay constant $f_a$) associated with an ALP can be unrelated~\cite{Dimopoulos:1979pp,Tye:1981zy,Zhitnitsky:1980tq,Dine:1981rt,Holdom:1982ex,Kaplan:1985dv,Srednicki:1985xd,Flynn:1987rs,Kamionkowski:1992mf,Berezhiani:2000gh,Hsu:2004mf,Hook:2014cda,Alonso-Alvarez:2018irt,Hook:2019qoh}. The range of ALP mass may span from sub-micro-eV~\cite{Kim:1979if,Shifman:1979if,Dine:1981rt,Zhitnitsky:1980tq,Turner:1989vc} to TeV scale and even beyond~\cite{Rubakov:1997vp,Fukuda:2015ana,Gherghetta:2016fhp,Dimopoulos:2016lvn,Chiang:2016eav,Gaillard:2018xgk,Gherghetta:2020ofz}. Thus, the search for the ALPs requires rather different strategies and facilities in experiments. The probe of ALP also depends on the various interactions between ALP and the SM matter. Here, we are interested in the ALP-electron coupling defined via the Lagrangian term
\begin{eqnarray}
\mathcal{L}_{\rm ALP}\supset c_{ae} {\partial_\mu a\over 2f_a} \overline{e} \gamma^\mu \gamma_5 e\;,
\end{eqnarray}
where $c_{ae}$ is a dimensionless constant. One usually defines another dimensionless ALP-electron coupling $g_{ae}=c_{ae}m_e/f_a$.
The amplitude of Compton scattering to ALP is
\begin{eqnarray}
\calm^{a}
&=&{c_{ae}\over 2f_a} {1\over \sqrt{2k^{\prime 0} V}}e^{ik'\cdot x}\overline{\psi_{p', s'}(x)} \cancel{k}' \gamma_5 \psi_{p, s}(x)\nonumber\\
&=&
{c_{ae}\over 2f_a} {e^{i(k'+q'-q)\cdot x}e^{-i\Phi}\over \sqrt{2^3 V^3 q^0 q^{\prime 0} k^{\prime 0}}} \overline{u(p', s')} \left[1-\frac{e \slashed{A}\slashed{k}  }{2 k \cdot p'}\right]\cancel{k}' \gamma_5 \left[1-\frac{e \slashed{k} \slashed{A} }{2 k \cdot p}\right] u(p,s)\;.
\end{eqnarray}
The individual amplitudes $\calm_i^a$ become
\bea
\calm_0^a
&=&
\overline{u_2} \cancel{k}'\gamma_5 u_1 +\overline{u_2}{e^2 a^2\over 2k\cdot p k\cdot p'} k\cdot k' \cancel{k}\gamma_5 u_1\;,
\\[3mm]
\calm_1^a
&=&
-\overline{u_2}\cancel{k}'\gamma_5
\frac{e \slashed{k} \slashed{a}_1 }{2 k \cdot p} u_1 - \overline{u_2}
\frac{e \slashed{a}_1 \slashed{k} }{2 k \cdot p'} \cancel{k}'\gamma_5 u_1\;,
\\[3mm]
\calm_2^a
&=&
-\overline{u_2}\cancel{k}'\gamma_5
\frac{e \slashed{k} \slashed{a}_2 }{2 k \cdot p} u_1 - \overline{u_2}
\frac{e \slashed{a}_2 \slashed{k} }{2 k \cdot p'} \cancel{k}'\gamma_5 u_1\;.
\ena
The spin averaged and squared amplitude becomes
\begin{eqnarray}
\sum_{i,j=0}^{2}
\widetilde{C}_i^{n} \big( \widetilde{C}_j^{n} \big)^\ast  \overline{ \calm_i^a \calm_j^{a\dag} }&=&-4m_e^2 m_\chi^2J_n^2(z)\nonumber\\
&+&e^2a^2{2m_e^2(1-u)^2\over u} [J_{n-1}^2(z)+J_{n+1}^2(z)-2J_n^2(z)]\;,
\end{eqnarray}
where $m_\chi=m_a$. The cross section of relevant Compton scattering is obtained by substituting $e^2$ in Eq.~(\ref{eq:xsec}) by $(c_{ae}/2f_a)^2$.

\section{Numerical results}
\label{sec:results}

In this section, we show the numerical results of laser induced Compton scattering to DP or ALP. We define an intensity quantity in the following calculation
\begin{eqnarray}
\eta\equiv {ea\over m_e}={e \varepsilon_0\over \omega_{\rm Lab} m_e}\;,
\end{eqnarray}
where $\omega_{\rm Lab}$ denotes the laser beam energy in the laboratory frame.
In the SLAC experiment, the electron beam has the energy of $E_{\rm Lab}=46.6$ GeV and collides an intense laser beam of green light with $\omega_{\rm Lab}=2.35$ eV. The scattering angle is $\theta_{\rm Lab}=17^\circ$. Given the green light energy, $\eta=1$ corresponds to $\varepsilon_0\simeq 10^{10}~{\rm V/cm}$. We next show the numerical results with the setup of the SLAC experiment for illustration. As stated before, this quantity $\eta$ characterizes the nonlinear effects of the multiphoton processes.
The quantity $\eta$ exhibits a maximal allowed value for a fixed $m_\chi$ and this threshold enhances with increasing $n$
\begin{eqnarray}
\eta^{\rm max}=\Big[\Big({2n\omega_{\rm Lab}(E_{\rm Lab}+\cos\theta_{\rm Lab}p_{\rm Lab})-m_\chi^2\over 2m_e m_\chi}\Big)^2-1\Big]^{1/2}\;,~~~p_{\rm Lab}=\sqrt{E_{\rm Lab}^2-m_e^2}\;.
\end{eqnarray}

\begin{figure}[ht]
\begin{center}
\includegraphics[width=0.48\textwidth]{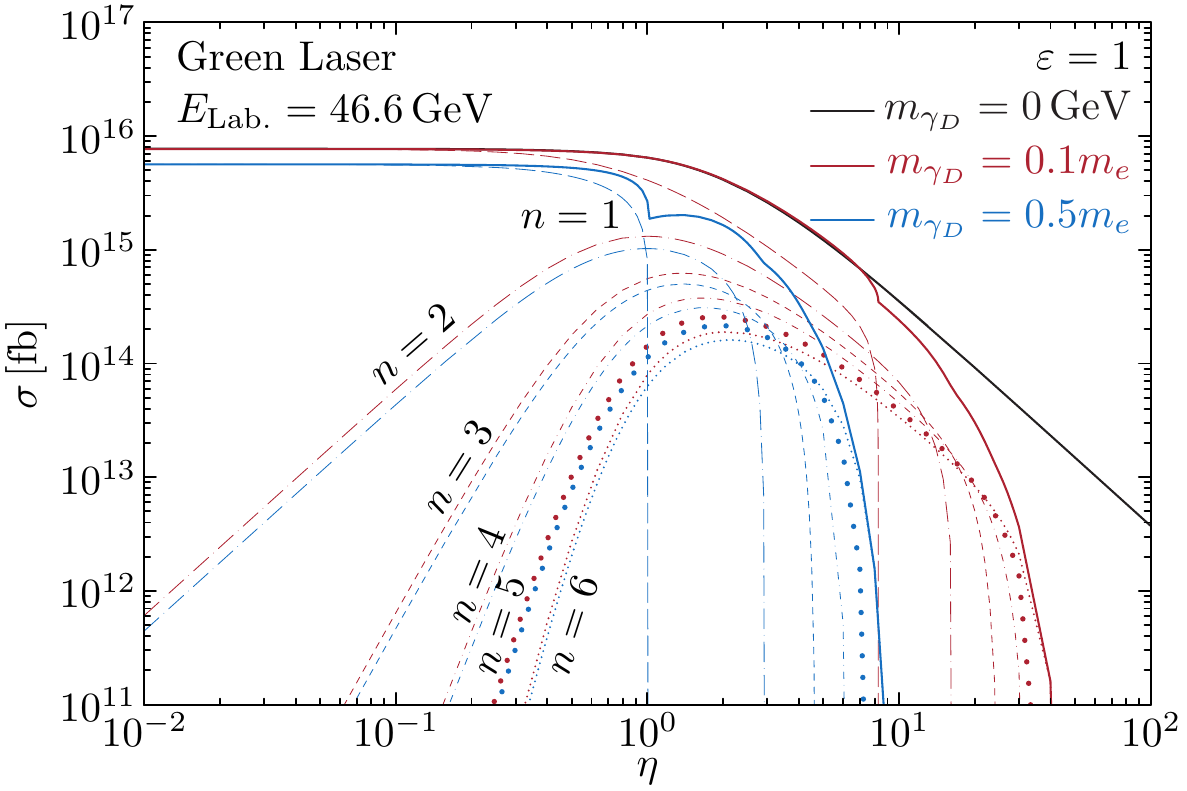}
\includegraphics[width=0.48\textwidth]{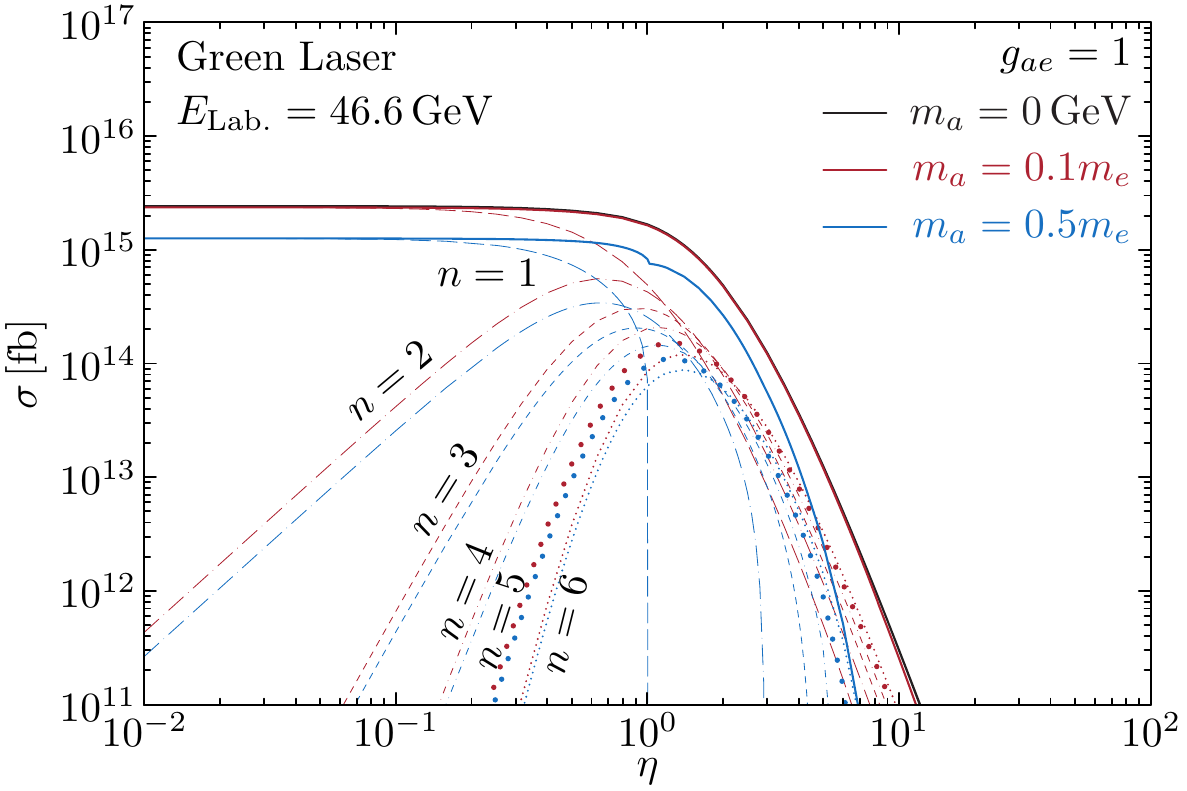}
\caption{The cross section of laser induced Compton scattering to DP with $\epsilon=1$ (left) or ALP with $g_{ae}=1$ (right), as a function of quantity $\eta$. The contributions from the absorption of $n=1$ up to $n=6$ photons are displayed. The summed values are shown in solid lines. The mass of new particle is assumed to be 0 (black), $0.1m_e$ (red) and $0.5m_e$ (blue).
}
\label{fig:DP:Eta}
\end{center}
\end{figure}

The cross sections of laser induced Compton scattering to DP with $\epsilon=1$ (left) or ALP with $g_{ae}=1$ (right), as a function of quantity $\eta$, are displayed in Fig.~\ref{fig:DP:Eta}. The contributions from the absorption of $n=1$ up to $n=6$ photons are shown. The summed values are shown in solid lines. The mass of new particle is assumed to be 0 (black), $0.1m_e$ (red) and $0.5m_e$ (blue). The black solid line indicates the total cross section of pure Compton scattering to a single photon.
As seen from Fig.~\ref{fig:DP:Eta}, the absorption of multiple laser photons leads to a series of edges of the cross section beyond the exact $n=1$ result. For a given number $n$, when $\eta$ exceeds unity and gets larger, more nonlinear effects contribute to the cross section and become important. Nevertheless, the nonlinearity decreases the individual cross section due to the increased effective mass $m_e^\ast$ in final states.

\begin{figure}[ht]
\begin{center}
\includegraphics[width=0.48\textwidth]{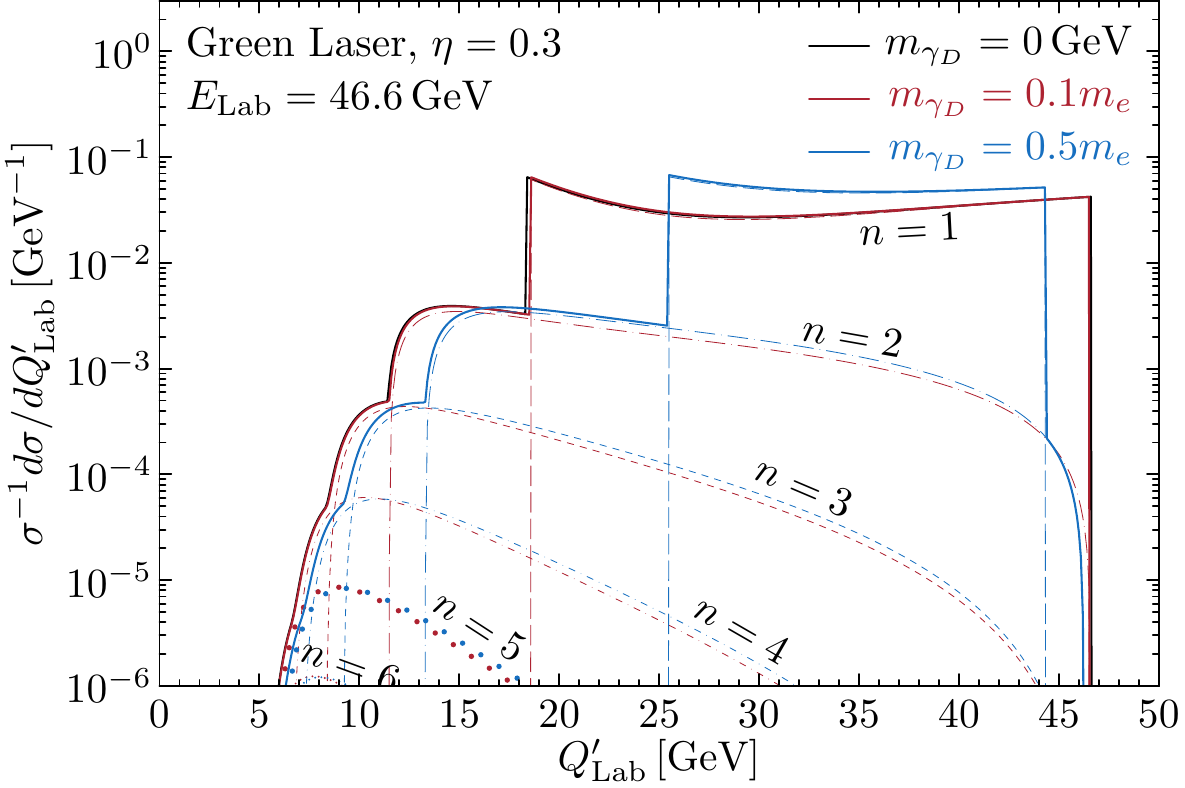}
\includegraphics[width=0.48\textwidth]{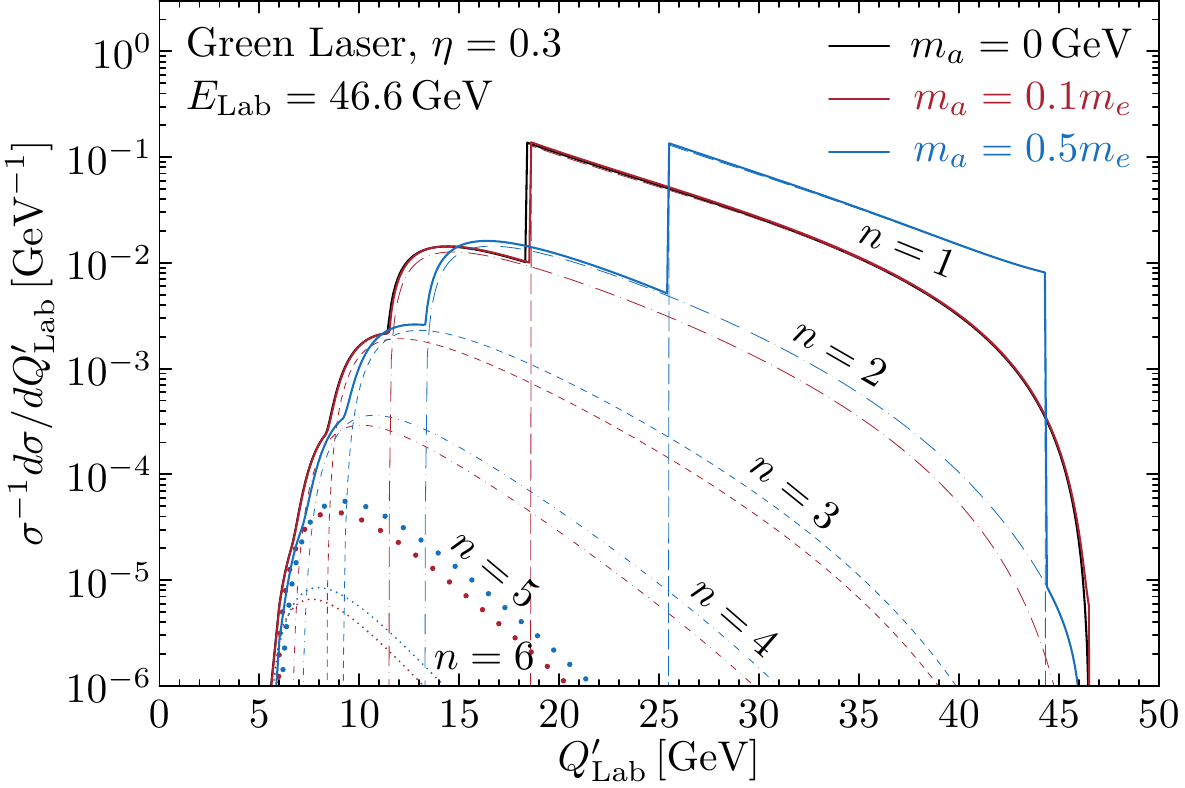}
\caption{The normalized distribution of the outgoing electron energy for DP (left) or ALP (right). The mass of new particle is assumed to be 0 (black), $0.1m_e$ (red) and $0.5m_e$ (blue). We take $\eta=0.3$ for illustration.
}
\label{fig:DP:EE}
\end{center}
\end{figure}

In Fig.~\ref{fig:DP:EE}, we show the energy distribution of the scattered electron in the laboratory frame $Q_{\rm Lab}'$. We take $\eta=0.3$ and $m_\chi=0$ (black), $0.1m_e$ (green) and $0.5m_e$ (cyan). Larger new particle mass shifts the outgoing electron to higher energy and results in narrower energy range.
As the number $n$ increases, the outgoing electron gains energy lower than the $n=1$ kinematic edge and the differential cross section decreases.

\begin{figure}[ht]
\begin{center}
\includegraphics[width=0.49\textwidth]{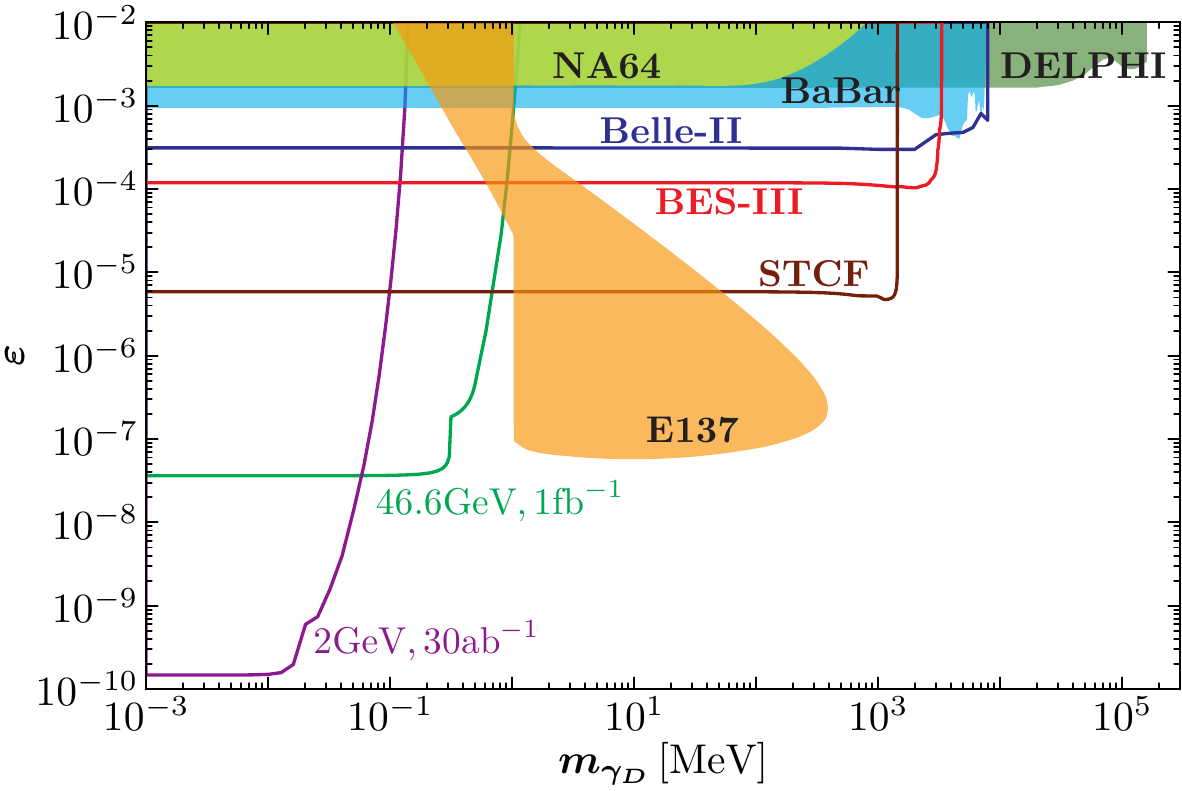}
\includegraphics[width=0.49\textwidth]{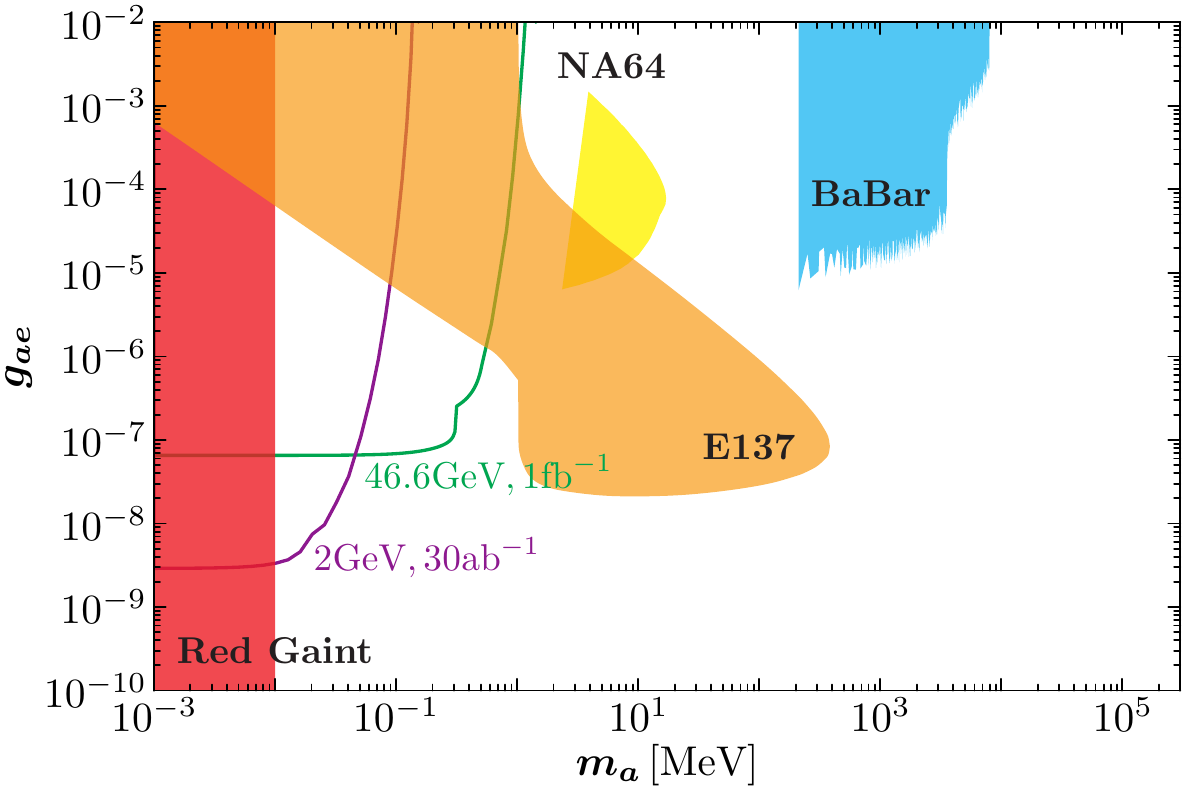}
\caption{Sensitivity of laser induced Compton scattering to DP kinetic mixing $\epsilon$ (left) and ALP-electron coupling $g_{ae}$ (right). We take $E_{\rm Lab}=46.6$ GeV with $\mathcal{L}=1~{\rm fb}^{-1}$ (green line) and $E_{\rm Lab}=2$ GeV with $\mathcal{L}=30~{\rm ab}^{-1}$ (purple line). The limits from other collider and beam dump experiments are also shown. For DP, we include the exclusion limits from BaBar (light blue region)~\cite{BaBar:2017tiz}, E137 (orange region)~\cite{Liu:2017htz}, NA64$\mu$ (light green region)~\cite{Andreev:2024yft} and DELPHI (green region) \cite{Ma:2022cto}, and the projection limits from Belle II ($20~{\rm fb}^{-1}$, blue line)~\cite{Belle-II:2018jsg}, BESIII ($17~{\rm fb}^{-1}$, red line)~\cite{Zhang:2019wnz} and STCF (2 GeV, $30~{\rm ab}^{-1}$, brown line)~\cite{Zhang:2019wnz}. For ALP, we include exclusion limits from BaBar (light blue region)~\cite{Bauer:2017ris,Bauer:2018uxu,Liu:2023bby}, E137 (orange region)~\cite{Liu:2017htz}, NA64 (yellow region)~\cite{NA64:2021aiq} and red giant (red region)~\cite{Viaux:2013lha, Capozzi:2020cbu}.
}
\label{fig:DP:Sen}
\end{center}
\end{figure}

Fig.~\ref{fig:DP:Sen} shows the sensitivity of laser induced Compton scattering to the kinetic mixing of DP $\epsilon$ (left) and the ALP-electron coupling $g_{ae}$ (right). In this figure, we take the laser energy of green light and $E_{\rm Lab}=46.6$ GeV or 2 GeV. We assume the observed number of events as 10 and various integrated luminosities ($\mathcal{L}=1~{\rm fb}^{-1}$ and $30~{\rm ab}^{-1}$) to get the sensitivity limits. The limits from other collider and beam dump experiments are also shown for comparison (as well as red giant limit for ALP~\footnote{For a particle with mass below the relevant plasma frequency of about $10\sim20~{\rm keV}$~\cite{Viaux:2013lha},
it can be emitted from the degenerate helium core near the red-giant branch. Here we choose the most conservative value, i.e., $m_a < 10~{\rm keV}$, for the red giant limit.}). One can see that the laser induced process provides a complementary search of invisible particles for $m_\chi<1$ MeV, compared with beam dump experiments.
Similar to other collider experiments that search for invisible particles using single-photon events with missing energy~\cite{Belle-II:2018jsg,Zhang:2019wnz}, we propose to search for single-electron events with missing energy. Although this signal is quite common at colliders, as evidenced by the discovery of the $W$ boson, the simulation of missing energy event production in laser experiments is beyond the scope of this theoretical work. For a rough estimation, we simply require the detection of 10 events. Given the argument of negligible SM background discussed below, this roughly corresponds to a
$3\sigma$ observation based on the significance formula~\cite{ParticleDataGroup:2024cfk}
\begin{eqnarray}
{S\over \sqrt{S+B}}\;,
\end{eqnarray}
where $S$ ($B$) denotes the number of signal (background) events.
A more detailed study of missing energy events in laser experiments is expected in future work.

Finally, we discuss the detection mechanism of the dark particles. For illustration, the light ALP may be detected through the coupling of the ALP to photons
\begin{eqnarray}
\mathcal{L}_{\rm ALP}\supset -{1\over 4}g_{a\gamma\gamma} a F^{\mu\nu} \tilde{F}_{\mu\nu}\;,
\end{eqnarray}
where $F_{\mu\nu}$ is the SM electromagnetic field strength and $\tilde{F}_{\mu\nu}$ is its Hodge dual. Then, the sensitivity bound in the right panel of Fig.~\ref{fig:DP:Sen} should be dependent on the combination of the ALP coupling to the electron and the ALP decay branching fraction to a pair of photons, i.e., $g_{ae}\sqrt{{\rm BR}(a\to \gamma\gamma)}$. However, for light ALP, the ALP-photon coupling is highly constrained as $g_{a\gamma\gamma}\lesssim 10^{-11}~{\rm GeV}^{-1}$~\cite{Ayala:2014pea,CAST:2017uph}. Based on the total decay width
\begin{eqnarray}
\Gamma(a\to \gamma\gamma)={g_{a\gamma\gamma}^2 m_a^3\over 64\pi}\;,
\end{eqnarray}
the ALP with mass less than 1 MeV has decay length larger than $10^{16}~{\rm m}$ and is quite long-lived.
Moreover, light DP can decay through the three-photon channel and two-neutrino channel with $\Gamma(\gamma_D\to 3\gamma)\sim \epsilon^2 m_{\gamma_D}^9/m_e^8$ and $\Gamma(\gamma_D\to \nu\overline{\nu})\sim \epsilon^2 m_{\gamma_D}^5/m_Z^4$. Both of them are highly suppressed. We presume the DP or ALP with mass less than 1 MeV is long-lived particle and flies out of the detector. Thus, our signal event is composed of one single electron and missing energy carried away by the DP or ALP. The single electron is the only detectable object in our final states. The different shapes of its energy distribution in Fig.~\ref{fig:DP:EE} can help us to distinguish the radiation of DP or ALP. This kind of detection signal is quite common in realistic collider experiments such as the LHC, characterized as mono-electron signature~\cite{Bai:2012xg,Ma:2024aoc,ATLAS:2014wra,CMS:2014fjm}.
The SM background then becomes
\begin{eqnarray}
e^- + {\rm laser} \to e^- + \nu +\overline{\nu}\;,
\end{eqnarray}
though an approximate four-fermion weak interaction mediated by the heavy SM gauge bosons $W^\pm$ or $Z$.
For illustration, the ratio of production amplitudes of DP and the SM processes is roughly given by
\begin{eqnarray}
e\epsilon : {g_Z^2 m_e^{\ast 2}\over m_Z^2} \approx \epsilon : 5\times 10^{-11}\;,
\end{eqnarray}
where $g_Z=e/(\sin\theta_W \cos\theta_W)$ with $\sin\theta_W$ ($\cos\theta_W$) being the sine (cosine) of the weak mixing angle $\theta_W$. The SM background should be taken into account when the reach of kinetic mixing $\epsilon$ is smaller than about $10^{-11}$. This is much lower than the reachable limit even for the more optimistic luminosity choice of $\mathcal{L}=30~{\rm ab}^{-1}$ as seen in Fig.~\ref{fig:DP:Sen}. Thus, it is safe for us to neglect the SM background here and simply take 10 observed events to estimate the sensitivity.

Alternatively, Ref.~\cite{Dillon:2018ouq} proposed to detect the ALP through the LSW setup in the regeneration region. This approach requires to extend the laser experiment to include a strong magnetic field which converts the ALP to a photon. The bound of the coupling product $g_{ae}g_{a\gamma\gamma}$ can reach as small as about $10^{-13}~{\rm GeV}^{-1}$ for $m_a\lesssim 10^{-4}~{\rm eV}$. This reachable limit is very likely excluded by the existing bounds. Nevertheless, using this approach, they claimed that there are opportunities to probe heavier ALP regime of $10~{\rm eV}<m_a<100~{\rm eV}$ outside other bounds. As this proposal relies on additional experimental setup, the practicality of relevant technology is beyond the scope of our study. We refer the readers to the original study in Ref.~\cite{Dillon:2018ouq} for further details on the envisaged proposal and the projected exclusion bounds.

\section{Conclusion}
\label{sec:Con}

The nonlinear Compton scattering was observed in the collision of
an ultra-relativistic electron beam with a laser pulse of an intense electromagnetic field. The precision measurement of the nonlinear Compton scattering shines the light on the studies of strong-field QED and nonlinear effects in QED.

In this work, we propose to produce new particles through the nonlinear Compton scattering in the presence of a high intense laser field. We take dark photon or axion-like particle as illustrative new invisible particle and assume the existence of their couplings to electron. A high-energy electron beam collides with an intense laser beam of a number of optical photons and radiates a massive particle. The Compton scattering cross section is calculated in strong-field QED. We find that
\begin{itemize}
\item The absorption of multiple laser photons leads to a series of edges of the cross section beyond the exact $n=1$ result. For a given number $n$, when $\eta$ exceeds unity and gets larger, more nonlinear effects contribute to the cross section.
\item As the number $n$ increases, the outgoing electron gains energy lower than the $n=1$ kinematic edge and the differential cross section decreases.
\item The laser induced process provides a complementary search of new invisible particles for $m_\chi<1$ MeV, compared with beam dump experiments.
Similar to other collider experiments that search for invisible particles using single-photon events with missing energy, we propose to search for single-electron events with missing energy. This approach requires more detailed experimental study in future work.
\end{itemize}

\acknowledgments

We would like to thank M.~Ouhammou, M.~Ouali and B.~Manaut for useful discussions. T.~L. is supported by the National Natural Science Foundation of China (Grant No. 12375096, 12035008, 11975129) and ``the Fundamental Research Funds for the Central Universities'', Nankai University (Grants No. 63196013). K. M. was supported by the Natural Science Basic Research Program of Shaanxi (Program No. 2023-JC-YB-041), and the Innovation Capability Support Program of Shaanxi (Program No. 2021KJXX-47).

\appendix

\section{Lorentz transformation}
\label{app:rotation}

In case that the incoming laser and electron beam are not colliding head-on, the further Lorentz rotations are necessary. We suppose the incoming momenta are defined as follows
\bea
k^\mu &=& \omega_{\rm Lab} (1,\, 0,\, 0,\, 1)\;,
\\[3mm]
p^\mu &=& E_{\rm Lab} (1,\, \sin\theta_{\rm Lab}\beta_{\rm Lab},\, 0,\, -\cos\theta_{\rm Lab}\beta_{\rm Lab})\;,
\ena
where $\pi-\theta_{\rm Lab}$ is the intersection angle between the laser and electron momenta.
Then we have
\bea
k\cdot p &=& E_{\rm Lab} \omega_{\rm Lab} (1+ \cos\theta_{\rm Lab}\beta_{\rm Lab}) = \frac{1}{2} (s - m_e^2)\;,
\\[3mm]
s &=& (k+p)^2 = m_e^2 + 2E_{\rm Lab} \omega_{\rm Lab} (1+ \cos\theta_{\rm Lab}\beta_{\rm Lab})\;,
\\[3mm]
s_n &=& (nk+p)^2 = m_e^2 + 2nE_{\rm Lab} \omega (1+ \cos\theta_{\rm Lab}\beta_{\rm Lab}) = n s - (n-1) m_e^2\;.
\ena
One can see that $s$ becomes slightly smaller,
and all the other quantities' expressions (in terms of $s$) do not change.
Hence, we only need to consider the effect of Lorentz transformations
on the outgoing photon energy. Firstly, the total momentum is
\bea
s^\mu_{\rm Lab}
&=&
(E_{\rm Lab} +\omega_{\rm Lab},\,
\sin\theta_{\rm Lab}\beta_{\rm Lab}E_{\rm Lab},\,
0,\,
\omega_{\rm Lab} - \cos\theta_{\rm Lab}\beta_{\rm Lab}E_{\rm Lab})\nonumber
\\[3mm]
&=&
(E_{\rm Lab} +\omega_{\rm Lab})(1,\, \sin\theta_s\beta_s,\, 0,\, \beta_s \cos\theta_s)\;,
\ena
where the velocity is given by
\bee
\gamma_s = \frac{E_{\rm Lab} +\omega_{\rm Lab}}{\sqrt{s} }
\quad,\quad
\beta_s = \sqrt{1-\frac{1}{ \gamma_s^2 }}
=
\frac{\sqrt{ (E_{\rm Lab} +\omega_{\rm Lab})^2 -s} }{ E_{\rm Lab} +\omega_{\rm Lab} }\;.
\ene
The sine and cosine of its polar angle are
\bea
\cos\theta_s
=
\frac{ \omega_{\rm Lab} - \cos\theta_{\rm Lab}p_{\rm Lab} }{ \beta_s (E_{\rm Lab} + \omega_{\rm Lab})  }\;,~~~
\sin\theta_s
=
\frac{ \sin\theta_{\rm Lab} p_{\rm Lab} }{ \beta_s (E_{\rm Lab} + \omega_{\rm Lab})  }\;.
\ena
The momentum of the outgoing photon in the laboratory frame can be obtained
with following Lorentz transformations
\bee
R_{s'_n} \;\longrightarrow\; R_{K'} \;\longrightarrow\; R_K \;\longrightarrow\; R_{\rm Lab}\;,
\ene
where the rest frame $R_K$ and $R_{K'}$ are defined by the total momentum such that
\bea
s^\mu_{K'}
&=&
\sqrt{s}(1,\, 0,\, 0,\, 0 )\;,
\\[3mm]
k^\mu_{K'} &=& \omega_{K'} (1,\, 0,\, 0,\, 1)\;,
\\[3mm]
p^\mu_{K'} &=& E_{K'} (1,\, 0,\, 0,\, -\beta_{K'})\;,
\ena
and
\bea
s^\mu_K
&=&
\sqrt{s}(1,\, 0,\, 0,\, 0 )\;,
\\[3mm]
k^\mu_K &=& \omega_K (1,\, \sin\theta_K,\, 0,\, \cos\theta_K)\;,
\\[3mm]
p^\mu_K &=& E_K (1,\, -\beta_K\sin\theta_K,\, 0,\, -\beta_K\cos\theta_K)\;.
\ena
It is clear they are related by only a rotation. As a result, we have
$\omega_{K'} =\omega_{K}$, $E_{K'} =E_{K} $ and $\beta_{K'} = \beta_{K}$.
One can easily find that
\bee
s^\mu_{\rm Lab}
=
\calr_{Y}(\theta_s)  \calb_{Z}(\beta_s) s^\mu_{K}
\ene
with
\bee
\calb_{Z}(\beta_s)
=
\left(
  \begin{array}{cccc}
    \gamma_s &  &  & \gamma_s \beta_s \\
     & 1 &  &  \\
     &  & 1 &  \\
    \gamma_s \beta_s &  &  & \gamma_s \\
  \end{array}
\right)
\quad,\quad
\calr_{Y}(\theta_s)
=
\left(
  \begin{array}{cccc}
    1 &  &  &  \\
     & \cos\theta_s &  & \sin\theta_s \\
     &  & 1 &  \\
     & -\sin\theta_s &  & \cos\theta_s \\
  \end{array}
\right)\;.
\ene
Now we need to calculate the momentum of the laser and electron
in the reference frame $R_{K}$ and $R_{K'}$.
For laser we have
\bea
k^\mu_{K}
&=&
\calb_{Z}(-\beta_s) \calr_{Y}(-\theta_s) k^\mu_{\rm Lab} \nonumber
\\[3mm]
&=&
\omega_{\rm Lab} \calb_{Z}(-\beta_s) \calr_{Y}(-\theta_s)(1,\, 0,\, 0,\, 1) \nonumber
\\[3mm]
&=&
\omega_{\rm Lab}  \calb_{Z}(-\beta_s) (1,\, -\sin\theta_s,\, 0,\, \cos\theta_s) \nonumber
\\[3mm]
&=&
\omega_{\rm Lab}
\Big[ \gamma_s(1-\beta_s\cos\theta_s),\, -\sin\theta_s,\, 0,\, \gamma_s(\cos\theta_s - \beta_s) \Big]\;.
\ena
Similarly, the momentum of the electron is given by
\bea
p^\mu_{K}
&=&
E_{\rm Lab} \calb_{Z}(-\beta_s) \calr_{Y}(-\theta_s)  (1,\, \beta_{\rm Lab}\sin\theta_{\rm Lab},\, 0,\, -\beta_{\rm Lab}\cos\theta_{\rm Lab}) \nonumber
\\[3mm]
&=&
E_{\rm Lab} \calb_{Z}(-\beta_s) \Big[
1,\,\beta_{\rm Lab} \sin(\theta_s+\theta_{\rm Lab}),\, 0,\, -\beta_{\rm Lab} \cos(\theta_s+\theta_{\rm Lab})
\Big]
\ena
and after boost it becomes
\bea
p^\mu_K =
E_{\rm Lab}
\left(
  \begin{array}{c}
    \gamma_s + \gamma_s\beta_s\beta_{\rm Lab} \cos(\theta_s+\theta_{\rm Lab}) \\
    \beta_{\rm Lab} \sin(\theta_s+\theta_{\rm Lab}) \\
    0 \\
    -\gamma_s\beta_{\rm Lab} \cos(\theta_s+\theta_{\rm Lab}) -  \gamma_s\beta_s \\
  \end{array}
\right)\;.
\ena
Hence we can obtain the following relation
\bea
\omega_K &=& \gamma_s\omega_{\rm Lab} ( 1 - \beta_s\cos\theta_s)\;,
\\[3mm]
E_K  &=& \gamma_s E_{\rm Lab} \Big[ 1+ \beta_s\beta_{\rm Lab} \cos(\theta_s+\theta_{\rm Lab}) \Big]\;.
\ena
One can easily find
\bea
\sin\theta_K &=& \frac{ -\omega_{\rm Lab}\sin\theta_s }{ \omega_K }
=
\frac{ -\sin\theta_s }{ \gamma_s ( 1 - \beta_s\cos\theta_s) }\;,
\\[3mm]
\cos\theta_K &=&
\frac{ \omega_{\rm Lab}\gamma_s(\cos\theta_s - \beta_s) }{ \omega_K }
=
\frac{ \cos\theta_s - \beta_s }{  1 - \beta_s\cos\theta_s }\;.
\ena
The rotation from $R_{K'} \to R_{K}$ is directly given by
\bee
\calr_{Y'}(\theta_K)
=
\left(
  \begin{array}{cccc}
    1 &  &  &  \\
     & \cos\theta_K &  & \sin\theta_K \\
     &  & 1 &  \\
     & -\sin\theta_K &  & \cos\theta_K \\
  \end{array}
\right)\;.
\ene

Now we have obtained all the elements for the Lorentz transformations.
For the first transformation $R_{s'_n} \to R_{K'}$, we define
\bee
\overline{\gamma}_{K'}
=
\frac{  \omega_{K'}  }{\omega^\ast}
= \overline{\gamma}_{K} =
\frac{  \omega_{K}  }{\omega^\ast}
=
\frac{ 2 n \gamma_s\omega_{\rm Lab} ( 1 - \beta_s\cos\theta_s) }{ \overline{\alpha'}^\ast  \sqrt{s_n'}  }\;,
\ene
where $\omega^\ast={\sqrt{s_n'}\over 2n}\Big(1-{m_e^{\ast 2}\over s_n'}\Big)$ and $\overline{\alpha'}^\ast=1-{m_e^{\ast 2}-m_\chi^2\over s_n'}$.
Then, we can obtain the boost factor of the laser as
\bee
\overline{\gamma}_K  = \sqrt{ \frac{ 1 + \beta }{ 1 - \beta } }
\quad \Rightarrow \quad
\beta = \frac{ \overline{\gamma}^2 - 1 }{ \overline{\gamma}^2 + 1 }
\quad , \quad
\gamma = \frac{ \overline{\gamma}^2 + 1 }{2 \overline{\gamma} }\;.
\ene
The momentum of the outgoing photon in the rest frame $R_{s'_n}$ is given by
\bee
k^{'\ast\mu} = \frac{1}{2}\sqrt{s_n'} ( \overline{\alpha'}^\ast,\,
-\overline{\beta}_n^{\ast} \boldsymbol{\pi}_n^\ast)\;,
\ene
where $\overline{\beta}_n^{\ast}=\lambda^{1/2}(s_n',m_e^{\ast 2},m_\chi^2)/s_n'$ and $\boldsymbol{\pi}_n^\ast=(\sin\theta_n^\ast \cos\phi_n^\ast,\sin\theta_n^\ast \sin\phi_n^\ast,\cos\theta_n^\ast)$.
Then, in the rest frame $R_{K'}$ it is
\bee
k^{'\mu}_{K'} = \frac{1}{2}\sqrt{s_n'} (
\gamma\overline{\alpha'}^\ast -  \gamma\beta\overline{\beta}_n^{\ast} \cos\theta^\ast ,\,
- \overline{\beta}_n^{\ast} \sin\theta^\ast\cos\phi^\ast ,\,
- \overline{\beta}_n^{\ast} \sin\theta^\ast\sin\phi^\ast,\,
 \gamma\beta\overline{\alpha'}^\ast - \gamma\overline{\beta}_n^{\ast} \cos\theta^\ast )\;.
\ene
After rotating to the reference frame $R_{K}$, we have
\bea
k^{'\mu}_{K}
&=&
\calr_{Y'}(\theta_K) k^{'\mu}_{K'}
\\[3mm]
&=&
\frac{1}{2}\sqrt{s_n'}
\left(
  \begin{array}{c}
    \gamma\overline{\alpha'}^\ast -  \gamma\beta\overline{\beta}_n^{\ast} \cos\theta^\ast  \\
    - \overline{\beta}_n^{\ast} \cos\theta_K \sin\theta^\ast\cos\phi^\ast
+
 \gamma \sin\theta_K \Big[ \beta\overline{\alpha'}^\ast - \overline{\beta}_n^{\ast} \cos\theta^\ast  \Big] \\
    - \overline{\beta}_n^{\ast} \sin\theta^\ast\sin\phi^\ast \\
    \overline{\beta}_n^{\ast} \sin\theta_K \sin\theta^\ast\cos\phi^\ast
+
 \gamma \cos\theta_K \Big[ \beta\overline{\alpha'}^\ast - \overline{\beta}_n^{\ast} \cos\theta^\ast  \Big] \\
  \end{array}
\right)\;.
\ena
For the transformation to the frame $R_{\rm Lab}$, we need a further rotation and a boost
\bea
&&k^{'\mu}_{\rm Lab}/\Big(\frac{1}{2}\sqrt{s_n'}\Big) = \calr_{Y}(\theta_s)  \calb_{Z}(\beta_s) k^{'\mu}_{K} /\Big(\frac{1}{2}\sqrt{s_n'}\Big)
\\[3mm]
&&=
\left(
  \begin{array}{c}
    \gamma_s\gamma \Big(\overline{\alpha'}^\ast -  \beta\overline{\beta}_n^{\ast} \cos\theta^\ast  \Big)
+
\gamma_s\beta_s\bigg[ \overline{\beta}_n^{\ast} \sin\theta_K \sin\theta^\ast\cos\phi^\ast
+
 \gamma \cos\theta_K \Big( \beta\overline{\alpha'}^\ast - \overline{\beta}_n^{\ast} \cos\theta^\ast  \Big) \bigg] \\
    - \overline{\beta}_n^{\ast} \cos\theta_K \sin\theta^\ast\cos\phi^\ast
+
 \gamma \sin\theta_K \Big[ \beta\overline{\alpha'}^\ast - \overline{\beta}_n^{\ast} \cos\theta^\ast  \Big] \\
    - \overline{\beta}_n^{\ast} \sin\theta^\ast\sin\phi^\ast \\
  \gamma_s\beta_s\gamma \Big(\overline{\alpha'}^\ast -  \beta_K\overline{\beta}_n^{\ast} \cos\theta^\ast  \Big)
+
\gamma_s\bigg[ \overline{\beta}_n^{\ast} \sin\theta_K \sin\theta^\ast\cos\phi^\ast
+
 \gamma \cos\theta_K \Big( \beta\overline{\alpha'}^\ast - \overline{\beta}_n^{\ast} \cos\theta^\ast  \Big) \bigg] \\
  \end{array}
\right)\;.\nonumber\\
\ena
Since the rotation does not change the energy of the outgoing photon, hence we have
\bea
\omega'_{\rm Lab}
&=&
\frac{1}{2}\sqrt{s_n'}\gamma_s\gamma \Big(\overline{\alpha'}^\ast -  \beta\overline{\beta}_n^{\ast} \cos\theta^\ast  \Big)\nonumber
\\[3mm]
&&
+
\frac{1}{2}\sqrt{s_n'}\gamma_s\beta_s\bigg\{ \overline{\beta}_n^{\ast} \sin\theta_K \sin\theta^\ast\cos\phi^\ast
+
 \gamma \cos\theta_K \Big[ \beta\overline{\alpha'}^\ast - \overline{\beta}_n^{\ast} \cos\theta^\ast  \Big] \bigg\}
\\[3mm]
&=&
\frac{1}{2}\sqrt{s_n'} \gamma_s\gamma  \overline{\alpha'}^\ast
\Big( 1+ \beta\beta_s \cos\theta_K  \Big)\nonumber
\\[3mm]
&&
+
\frac{1}{2}\sqrt{s_n'} \gamma_s \overline{\beta}_n^{\ast}  \Big[
\beta_s \sin\theta_K \sin\theta^\ast\cos\phi^\ast
- \gamma \Big( \beta  + \beta_s  \cos\theta_K \Big)  \cos\theta^\ast
\Big]
\ena
which can be abbreviated as
\bee
\omega'_{\rm Lab}
=
\frac{1}{2}\sqrt{s_n'} \Big[
\xi_2 \sin\theta^\ast - \xi_1 \cos\theta^\ast  + C\Big]
=
\frac{1}{2}\sqrt{s_n'} \Big[
C - \rho_\xi \cos(\theta^\ast + \theta_\xi)
\Big]
\ene
with
\bea
C &=& \gamma_s\gamma  \overline{\alpha'}^\ast
\Big( 1+ \beta\beta_s \cos\theta_K  \Big)\;,
\\[3mm]
\xi_1 &=&
\gamma_s \overline{\beta}_n^{\ast}
\gamma \Big( \beta  + \beta_s  \cos\theta_K \Big) \;,
\\[3mm]
\xi_2 &=&
\gamma_s \overline{\beta}_n^{\ast} \beta_s \sin\theta_K \cos\phi^\ast\;,
\\[3mm]
\rho_\xi &=& \sqrt{\xi_1^2 + \xi^2_2}\;,
\\[3mm]
\cos\theta_\xi &=& \xi_1/\rho_\xi\;,
\\[3mm]
\sin\theta_\xi &=& \xi_2/\rho_\xi\;.
\ena
Since $\theta^\ast$ can take the full range values, the maximum and
minimum values for given $\phi^\ast$ of the $\omega'$ are given as
\bea
\omega'_{\rm Lab,\, Max }(\phi^\ast)
&=&
\frac{1}{2}\sqrt{s_n'} \big( C + \rho_\xi  \big)\;,
\\[3mm]
\omega'_{\rm Lab,\, Min }(\phi^\ast)
&=&
\frac{1}{2}\sqrt{s_n'} \big( C - \rho_\xi  \big)\;.
\ena
It is clear that when $\cos\phi^\ast = 1$, they take the extreme values
\bea
\omega'_{\rm Lab,\, Max }
&=&
\frac{1}{2}\sqrt{s_n'} \big[ C + \rho_\xi( \phi^\ast = 0 )  \big]\;,
\\[3mm]
\omega'_{\rm Lab,\, Min }
&=&
\frac{1}{2}\sqrt{s_n'} \big[ C - \rho_\xi( \phi^\ast = 0 )  \big]\;.
\ena
We find that
\bee
\theta^\ast
=
\arccos \frac{1}{ \rho_\xi }  \left( C - \frac{2\omega'_{\rm Lab } }{ \sqrt{s_n'} }   \right)
-
\theta_\xi\;.
\ene
Now we can use the following equation
\bee
u= \frac{ 2 }{ \alpha'^\ast - \overline{\beta'}^\ast \cos\theta^\ast}
\ene
to calculate the invariant $u$, and hence the variable $z$.
The differentials are given as
\bea
d \omega'_{\rm Lab}
&=&
\frac{1}{2}\sqrt{s_n'}\rho_\xi   \sin(\theta^\ast + \theta_\xi) d\theta^{\ast}
=
\frac{1}{2}\sqrt{s_n'} \rho_\xi   \frac{\sin(\theta^\ast + \theta_\xi)}{\sin\theta^\ast}d\cos\theta^{\ast}\;,
\\[3mm]
d\cos\theta^{\ast}
&=&
\frac{ 2 \sin\theta^\ast d \omega'_{\rm Lab}   }{ \sqrt{s_n'} \rho_\xi \sin(\theta^\ast + \theta_\xi) }
=
\frac{ 2 d \omega'_{\rm Lab}   }{ \sqrt{s_n'} \rho_\xi (\cos\theta_\xi + \cot\theta^\ast\sin\theta_\xi) }\;.
\ena
Finally, the cross section in the laboratory frame is given by
\bea
\sigma
&=& {1\over 2\rho_\omega} \frac{e^2}{32\pi^2 Q_{\rm Lab} }\sum_{n=0}^{\infty} \nonumber
\\[3mm]
&&
\int_{ \omega'_{\rm Lab,\, Min } }^{ \omega'_{\rm Lab,\, Max } } d \omega'_{\rm Lab} \int_0^{2\pi}
d\phi^\ast \frac{ \overline{\beta}_n^{\ast}    }
{ \sqrt{s_n'} \rho_\xi (\cos\theta_\xi + \cot\theta^\ast\sin\theta_\xi) }
\sum_{i,j=0}^{2}
\widetilde{C}_i^{n} \big( \widetilde{C}_j^{n} \big)^\dag
\overline{\calm_i \calm_j^\dag}\;,
\ena
where $Q_{\rm Lab}$ is the energy of incoming electron's $q$ momentum in the laboratory frame.
We can also easily obtain the energy of the outgoing electron $Q_{\rm Lab}'$ in the laboratory frame using energy conservation
\begin{eqnarray}
n\omega_{\rm Lab}+Q_{\rm Lab}=n\omega_{\rm Lab} + E_{\rm Lab} + {e^2 a^2\over 2E_{\rm Lab}(1+\cos\theta_{\rm Lab}\beta_{\rm Lab} )} = \omega_{\rm Lab}' + Q_{\rm Lab}'\;.
\end{eqnarray}

\bibliography{refs}

\end{document}